# Could John Ellard Gore have pre-empted the Hertzsprung-Russell diagram?

**Jeremy Shears**

## Abstract

John Ellard Gore (1845–1910) was an Irish amateur astronomer, prolific science writer and civil engineer. He was an inaugural member of the British Astronomical Association when it was founded in 1890 and was invited to become the first Director of its Variable Star Section. Through careful reasoning, Gore reached remarkably modern conclusions about the nature of stars, including their sizes, distances and luminosities. This paper explores whether he might have been able to produce an early form of what is now known as the Hertzsprung–Russell (H–R) diagram.

## Introduction

John Ellard Gore (1845–1910; Figure 1) was an Irish amateur astronomer, prolific science writer and civil engineer whose work helped to popularise astronomy during the late nineteenth and early twentieth centuries.

Born in Athlone, Ireland, Gore studied engineering at Trinity College Dublin before joining the Indian Public Works Department, where he spent many years working on the Sirhind Canal irrigation project in the Punjab. While living in India, he developed a keen interest in observational astronomy, using relatively modest telescopes to study variable stars, double stars and deep-sky objects. He discovered several variable stars, including W Cygni in 1884 and U Orionis in 1885.

After returning to Ireland and later settling in Dublin, Gore became one of the leading amateur astronomers of his day. He was a member of the Liverpool Astronomical Society, where he directed its Variable Star Section, and was an inaugural member of the British Astronomical Association when it was founded in 1890. He was invited to become the Association's first Director of the Variable Star Section. His observations of variable stars and binary star systems contributed to the growing understanding of stellar astronomy. He also investigated stellar distances, the nature of binary systems, and the possibility of "dark", or invisible, stellar companions. In addition, he speculated on the scale of the Universe and even entertained the possibility of multiple universes.

Among Gore's most original contributions were his attempts to determine the intrinsic brightness (luminosity) and physical sizes of stars at a time when reliable measurements of stellar distances were available for only a handful of objects. Much of what is now regarded as standard astrophysics still lay decades in the future, yet Gore used careful reasoning to reach conclusions that were remarkably prescient.

This paper examines whether Gore might have been able to produce an early form of the Hertzsprung–Russell (H–R) diagram from the data he had assembled on stellar luminosities and spectral classifications.

A review of Gore's life and astronomical contributions was presented by the present author in the April 2013 edition of this Journal (1).

**Gore's *immensity and minuteness* of stars**

In the late nineteenth century, accurate stellar parallaxes had been measured for only a handful of stars, leaving the distances to most stars unknown. Gore combined the best available parallax measurements with apparent magnitudes to estimate the intrinsic brightness of stars.

He recognised that:

> a star that appears faint may simply be very distant;
>
> once a star's distance is known, its intrinsic luminosity can be calculated;
>
> some stars must radiate hundreds or even thousands of times more light than the Sun.

This represented an important conceptual advance, since many astronomers of the period still regarded stars as being broadly similar to the Sun.

Gore became particularly interested in bright stars such as Arcturus and Capella. Using the best parallax measurements then available, he concluded that these stars possessed enormous intrinsic luminosities. As there was no accepted mechanism by which stars could be vastly hotter than the Sun, Gore argued that many highly luminous stars must instead be much larger. Although his numerical estimates were inevitably uncertain, he used straightforward calculations based on parallax and apparent brightness to estimate the relative sizes of a number of stars (2). In the case of Arcturus, he concluded it was a 'giant star' about 100 times the diameter of the Sun. Similarly, he estimated that Capella was a 'giant' with a diameter 18 times that of the Sun. Present day measurements place the diameters of Arcturus and Capella at 25.7 and 12.2 solar diameters respectively.

Another of Gore's notable investigations concerned the binary system Sirius. Although the faint companion, Sirius B, had been discovered visually, its physical nature remained a mystery. Gore calculated that if Sirius B were comparable in mass to the Sun yet emitted so little light, its density would have to be extraordinarily high—around 44,000 times the density of water. He regarded such a result as physically impossible (3) and therefore suggested that the companion was instead a large but intrinsically faint object.

Ironically, Gore's calculation was essentially correct. The extraordinary density that he considered impossible is now recognised as the defining characteristic of a white dwarf. Modern observations have shown that Sirius B is an Earth-sized star with a mass comparable to that of the Sun, making it one of the densest forms of matter known.

Gore's rejection of this extraordinary density is frequently cited as an example of how close nineteenth-century astronomers came to anticipating the physics of white dwarfs. He reached similar conclusions regarding the companions of Procyon and 40 Eridani, both of which are now known to be white dwarfs.

Gore was fascinated by the diversity of stars and by the scale of the Universe. In books such as *The Visible Universe* (1893) and *Studies in Astronomy* (1904), he emphasised that stars are not uniform objects but differ enormously in luminosity, size, colour, temperature and, as we would now describe it, evolutionary state.

This was a remarkably modern perspective at a time when no accepted theory of stellar evolution yet existed.

Although Gore lacked many of the tools available to twentieth-century astrophysicists — including modern spectroscopy, accurate parallaxes and, of course, the Hertzsprung–Russell diagram — his physical reasoning consistently pointed in the right direction.

Before considering whether Gore possessed sufficient information to construct his own H–R diagram ahead of Hertzsprung and Russell, it is useful first to examine how the diagram itself emerged.

**The emergence of the H-R diagram**

The Hertzsprung–Russell (H–R) diagram is a scatter plot showing the relationship between a star's intrinsic luminosity (or absolute magnitude) and its spectral classification (or effective temperature). Like many major scientific advances, the concept emerged independently through the work of Ejnar Hertzsprung (1873–1967) and Henry Norris Russell (1877–1957), and it represented a major step towards an understanding of stellar evolution.

It is important to appreciate that the H–R diagram did not appear fully formed. Rather, it evolved gradually between 1905 and 1913 as increasingly reliable data on stellar spectra and distances became available. Readers wishing to explore this development in greater detail are referred to David DeVorkin's classic review, *Stellar Evolution and the Origin of the Hertzsprung–Russell Diagram* (4). A brief summary will suffice here.

Hertzsprung's interest in astronomy stemmed in part from his father, who was an amateur astronomer. Beginning in 1905, Hertzsprung analysed stars that had been spectrally classified by Antonia Maury at the Harvard College Observatory. He demonstrated that some stars were intrinsically more luminous than others of the same spectral type. His paper included a graph relating apparent brightness to colour, using apparent magnitudes as a statistical proxy because reliable distances were then available for only a few stars. At the time, Hertzsprung was working as a photochemist, and his paper appeared in a relatively obscure German journal devoted to photographic science. Consequently, it attracted little immediate attention from astronomers (5) (6).

Hertzsprung subsequently extended his work using the small number of stars with measured parallaxes, producing plots that more directly related intrinsic brightness to stellar colour. Although these early papers contained sufficient tabulated data from which an H–R-type diagram could have been constructed—for example, for the Hyades, whose members he assumed to lie at approximately the same distance—he did not publish such a diagram until 1911 (7). Nielsen has suggested that Hertzsprung had in fact drawn a similar diagram as early as 1908 but withheld it from publication because of concerns about instrumental uncertainties (8).

In 1910, Hertzsprung's assistant, Hans Otto Rosenberg (1879–1940), produced a comparable diagram for the Pleiades (9), plotting spectral type against apparent magnitude for stars within the cluster, whose common distance made such a comparison meaningful. Although more accurately described as an early cluster colour–magnitude diagram, it is widely recognised as one of the earliest examples of what would become the H–R diagram.

Meanwhile, Russell was independently pursuing similar ideas between 1909 and 1913. Using greatly improved stellar parallaxes together with the Harvard spectral classification developed by Annie Jump Cannon, he plotted absolute magnitude against spectral type for a large sample of nearby stars.

Russell appears to have presented an early version of his diagram at the June 1913 meeting of the Royal Astronomical Society in London (10). Although no copy of the diagram shown on that occasion has survived, we know that during his visit he met J. Norman Lockyer (1836–1920), who was developing his own ideas on stellar evolution. Russell evidently showed Lockyer the tabulation reproduced here as Figure 2, a copy of which was clipped to a note that Lockyer sent to Russell during the visit (11). Although presented as a table rather than a graph, its essential features are unmistakable: the distribution of stars by spectral class and absolute magnitude foreshadows the structure of Russell's later diagram. Whether this table, or a graphical equivalent, was also shown during the Royal Astronomical Society meeting is not known.

What we do know is that Russell presented a fully graphical version of the diagram six months later, at a meeting in Atlanta, Georgia, in December 1913. The following year it appeared in print simultaneously in both *Nature* (12) and *Popular Astronomy* (13).

Russell's published diagram (Figure 3) clearly revealed two distinct stellar populations. The first was the now-familiar diagonal band extending from the upper left to the lower right, later named the main sequence. Above this lay the giant stars, occupying a distinct region of the diagram. A single star also appeared in the lower left. Russell believed this to represent an incorrect spectral classification; we now recognise it as a white dwarf (14).

In the following years, the diagram became better defined as new and more accurate data became available about luminosities and spectral type. On the other hand, its interpretation, especially regarding stellar evolution, has evolved considerably as our understanding on the internal constitution of the stars became better understood.

**Could Gore have produced an H-R diagram?**

Most historians of astronomy regard Hertzsprung's work (together with Rosenberg's cluster diagram) as the earliest example of an H–R-type diagram, while acknowledging that Russell's 1913 diagram established the form that subsequently became standard. Nevertheless, Gore's work raises an intriguing historical question: did he possess sufficient information to construct such a diagram several years earlier?

His published work suggests that he had assembled many of the essential ingredients. In particular, Gore:

- calculated intrinsic stellar luminosities from parallaxes and apparent magnitudes;
- recognised the enormous range in stellar luminosities and inferred corresponding differences in stellar size;
- discussed giant and intrinsically faint stars in physical, rather than purely descriptive, terms;
- speculated on stellar evolution;
- was familiar with the emerging Harvard spectral classification system; and
- was interested in comparing stars as physical objects rather than merely cataloguing them.

In his paper *On the Relative Brightness of Stars*, published in *Monthly Notices* in January 1905 (15), Gore tabulated data for 42 stars, including their luminosities relative to the Sun together with their spectral classifications, where available, obtained from Harvard College Observatory. His luminosity estimates were derived from apparent magnitudes—principally from the Harvard Photometry—combined with the best stellar parallaxes then available.

Taken together, these data contained many of the ingredients required to construct what we would now recognise as an H–R diagram.

There is, however, no evidence in Gore's published papers or books that he ever combined stellar luminosity and spectral classification in a graphical scatter plot. It is precisely this combination that distinguishes the H–R diagram from a numerical catalogue.

Indeed, Gore's preferred method of presenting data was through carefully compiled tables accompanied by descriptive discussion, rather than through graphical or statistical analysis. In this he was far from unusual. Throughout the nineteenth century, astronomical research relied overwhelmingly on numerical tables, whether in star catalogues, planetary ephemerides or eclipse predictions. Although graphical methods became increasingly common during the latter part of the century, they had not yet become standard practice in astronomical publications.

The same preference for tables extended beyond astronomy. Many Victorian statisticians regarded numerical tables as the primary scientific record, believing that the data should speak for themselves. Some even viewed graphical representations with suspicion, fearing that they encouraged subjective interpretation. Florence Nightingale was a notable exception. Her celebrated visualisations of mortality during the Crimean War demonstrated the persuasive power of graphical presentation and ultimately influenced hospital practice, illustrating an important milestone in the history of data visualisation (16).

The gradual adoption of graphical methods in astronomy reflected both changing scientific practice and practical considerations. As statistics and physics exerted increasing influence on astronomical research, and as advances in printing technology made the reproduction of figures more economical, graphs became more common. Solar activity plots and variable-star light curves, for example, were appearing with increasing frequency by the turn of the twentieth century.

Even so, many astronomical papers continued to consist entirely of numerical tables, and Gore remained firmly within that tradition. His reports on variable stars generally presented sequences of brightness estimates by date rather than plotted light curves, with relatively little attempt at graphical or statistical analysis.

Viewed in this context, Gore's failure to construct an H–R diagram appears less surprising. It was not simply that he lacked an insight available to Hertzsprung or Russell; rather, he was working within an established tradition in which numerical tabulation, rather than graphical representation, was the normal means of organising and interpreting astronomical data.

**The “Gore H-R diagram”**

Intrigued by the data presented in Gore's 1905 *Monthly Notices* paper, particularly his tabulated values for relative stellar luminosity and the accompanying Harvard spectral classifications, I attempted to reconstruct the diagram that Gore himself might have produced.

Following Russell's approach, I plotted the logarithm (base 10) of Gore's luminosities, expressed relative to the Sun, against the spectral classifications that he listed. Stars assigned to spectral class H, an early Harvard category denoting spectra dominated by strong hydrogen emission lines and later abandoned, were omitted from the analysis.

The resulting plot, shown in Figure 4, is strikingly reminiscent of Russell's 1914 diagram, despite being based on a much smaller sample of stars. A clear diagonal band extends from the upper left to the lower right, corresponding to what is now recognised as the main sequence. As in Russell's original figure, guide lines have been added simply to aid the eye.

Russell's diagram also showed a distinct population of giant stars lying above the main sequence. A similar pattern emerges from Gore's data. The stars that Gore had already recognised as giants—Arcturus, Capella, Antares and Aldebaran—occupy the upper-right region of the plot, separated from the principal sequence in much the same way as in Russell's diagram.

Two stars appear as conspicuous outliers in the lower-left corner of the reconstructed diagram. We now know that both result from incorrect or incomplete spectral classifications available at the time, rather than from any deficiency in Gore's luminosity calculations.

The upper of the two is Groombridge 1830 (Argelander's Star), a high-proper-motion star approximately 30 light-years distant. Gore listed it as spectral type A?, the question mark indicating uncertainty in the classification. It is now known to be a K-type main-sequence star. Had its modern classification been available, the star would have fallen naturally onto the main sequence.

The second outlier is $v^1$ Draconis (Philip, this is Greek letter Nu[1]). This chemically peculiar star exhibits unusually strong metallic absorption lines, giving rise to the complex modern spectral designation kA3hF0mF0. This notation indicates that the calcium K line corresponds to type A3, while both the hydrogen lines and the metallic spectrum correspond to type F0 lines (17). Its unusual spectrum explains why it occupies an anomalous position in the reconstructed diagram.

Although based on a rather small selection of stars and on the limited observational data available in 1905, the reconstructed "Gore H–R diagram" nevertheless reveals the principal features that later made Russell's diagram so significant. It suggests that Gore had assembled sufficient observational material to recognise the relationship between stellar luminosity and spectral class, had he chosen to present his data graphically.

**Closing remarks**

John Ellard Gore recognised that stars differ fundamentally from one another and that some must be vastly larger and more luminous than the Sun. He also came remarkably close to recognising the existence of stars of extraordinarily high density. Although he rejected the implications of his own calculations, we now recognise these objects as white dwarfs.

The Hertzsprung–Russell diagram has become one of the cornerstones of modern astrophysics, making its origins an important chapter in the history of astronomy. Gore's 1905 analysis demonstrates that he had already assembled much of the observational material needed to reveal the relationship between stellar luminosity and spectral class. Had he chosen to plot those quantities graphically, he might have recognised the characteristic diagonal band now known as the main sequence, together with a distinct grouping of giant stars.

Of course, so far as we know, Gore never took the crucial step of presenting his data in graphical form, and it was Hertzsprung and, above all, Russell who transformed the available observational evidence into a diagram that immediately revealed the underlying structure of the stellar population.

Yet the timing is itself revealing. During the opening years of the twentieth century, increasingly reliable measurements of stellar luminosities, parallaxes and spectral classifications were becoming available to astronomers. The emergence of the H–R diagram was therefore not an isolated event but the culmination of a gradual accumulation of observational evidence. In that sense, there was indeed something "in the air".

This does nothing to diminish the pioneering achievements of Hertzsprung and Russell. Rather, it highlights the remarkable insight of John Ellard Gore, who, working largely as an amateur astronomer, anticipated many of the physical ideas that would soon underpin modern stellar astrophysics. It also reminds us that scientific advances often depend not only on acquiring new observations but on finding new ways to organise and interpret existing data.

Gore was killed in a road traffic accident while crossing a street in Dublin in 1910, aged 65. It is intriguing to speculate how he might have responded to Russell's celebrated diagram, first presented just three years later, had he lived to see it.

**Acknowledgements**

The paper was motivated by questions and comments after a presentation by the author on Gore at the 2026 BAA Historical Section Meeting.

I am grateful to Damian Peach for kindly preparing the colourised image of Gore shown as Figure 1.

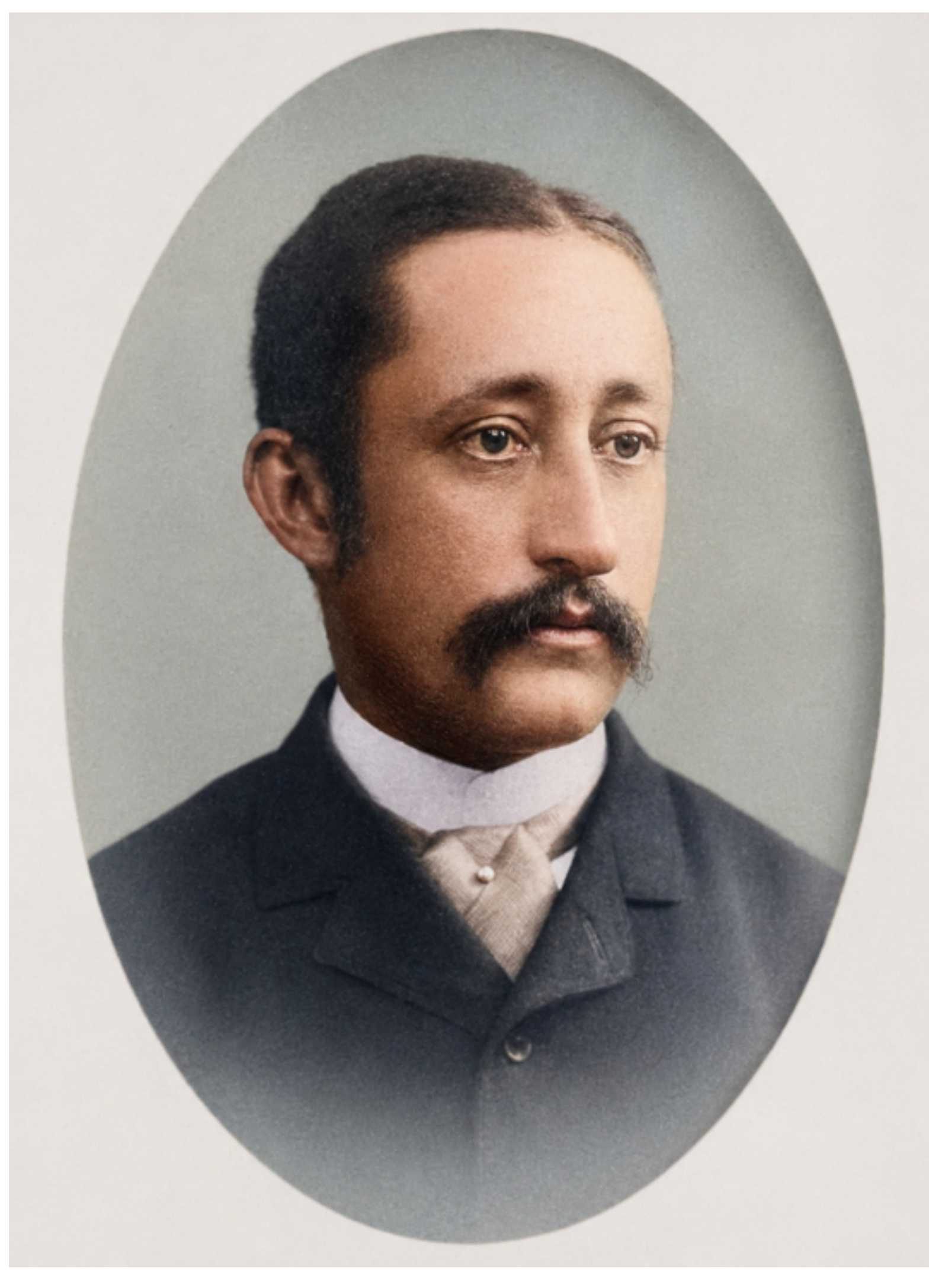

**Figure 1: John Ellard Gore**

The photograph appears to be from around 1875. It has been colourised by Damian Peach

| | O | B | A | F | G | K | M | N | R |
|---|---|---|---|---|---|---|---|---|---|
| | Oa-Oe | B0-B8 | B8-A3 | A3-F5 | F5-G5 | G5-K3 | K5-M5 | Na-Nb | Ra-R8 |
| -6.0 | | 2 | 1 | | | | | | |
| -4.5 | 1 | 5 | 2 | 2 | 1 | 1 | 1 | | |
| -3.0 | | 5 | 5 | 1 | 1 | 5 | 3 | | |
| -1.5 | 3 | 29 | 29 | 4 | 7 | 17 | 7 | 1 | |
| 0.0 | 5 | 74 | 83 | 20 | 18 | 63 | 24 | | |
| +1.5 | 2 | 122 | 201 | 31 | 40 | 167 | 49 | 7 | 1 |
| +3.0 | 2 | 140 | 522 | 89 | 72 | 418 | 120 | 11 | 2 |
| +4.5 | 5 | 58 | 809 | 271 | 199 | 817 | 189 | 10 | 4 |
| +6.0 | 3 | 50 | 855 | 469 | 448 | 1123 | 196 | 11 | 3 |
| +7.5 | 1 | 25 | 465 | 398 | 806 | 844 | 92 | 7 | 3 |
| +9.0 | | 7 | 280 | 140 | 726 | 566 | 26 | 1 | 2 |
| +10.5 | | | 68 | 16 | 277 | 331 | 9 | | |
| +12.0 | | | 7 | 6 | 47 | 115 | 12 | | |
| +13.5 | | | | | 5 | 16 | 7 | | |
| +15.0 | | | | | 1 | 4 | 2 | | |
| +16.5 | | | | | 1 | | 2 | | |
| +18.0 | | | 1 | | | | | | |

**Figure 2: Unpublished Russell table, June 1913** (Courtesy of Princeton University Library)

Spectral class on shown horizontally, absolute magnitude vertically. The digits in the table represent the number of stars in each magnitude/spectral class bucket.

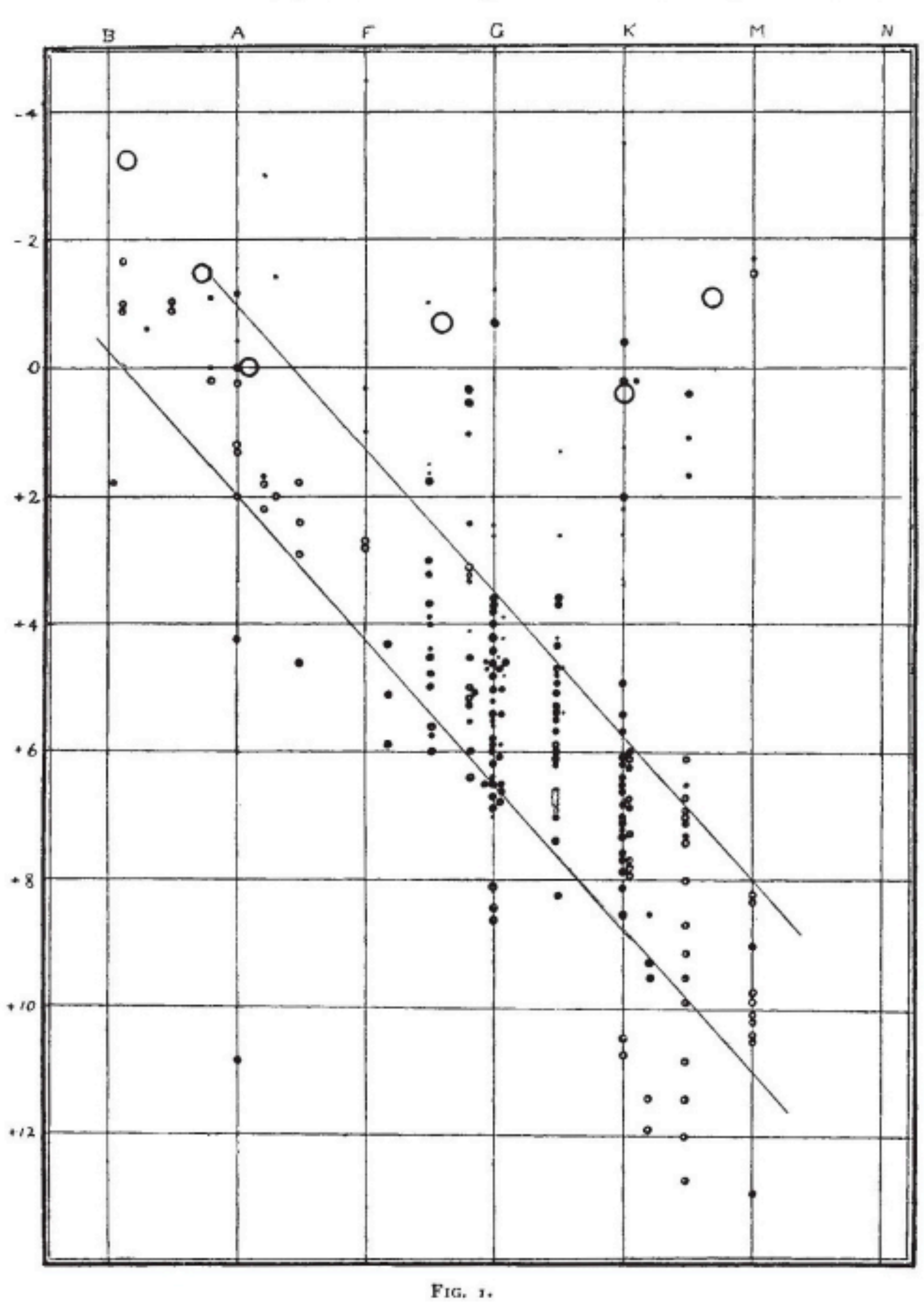

**Figure 3: The first printed Russell diagram from Russell's 1914 *Nature* paper**

Spectral class on x-axis, absolute magnitude on y-axis. Simultaneously published in 1914 in *Nature* and *Popular Astronomy* - see text.

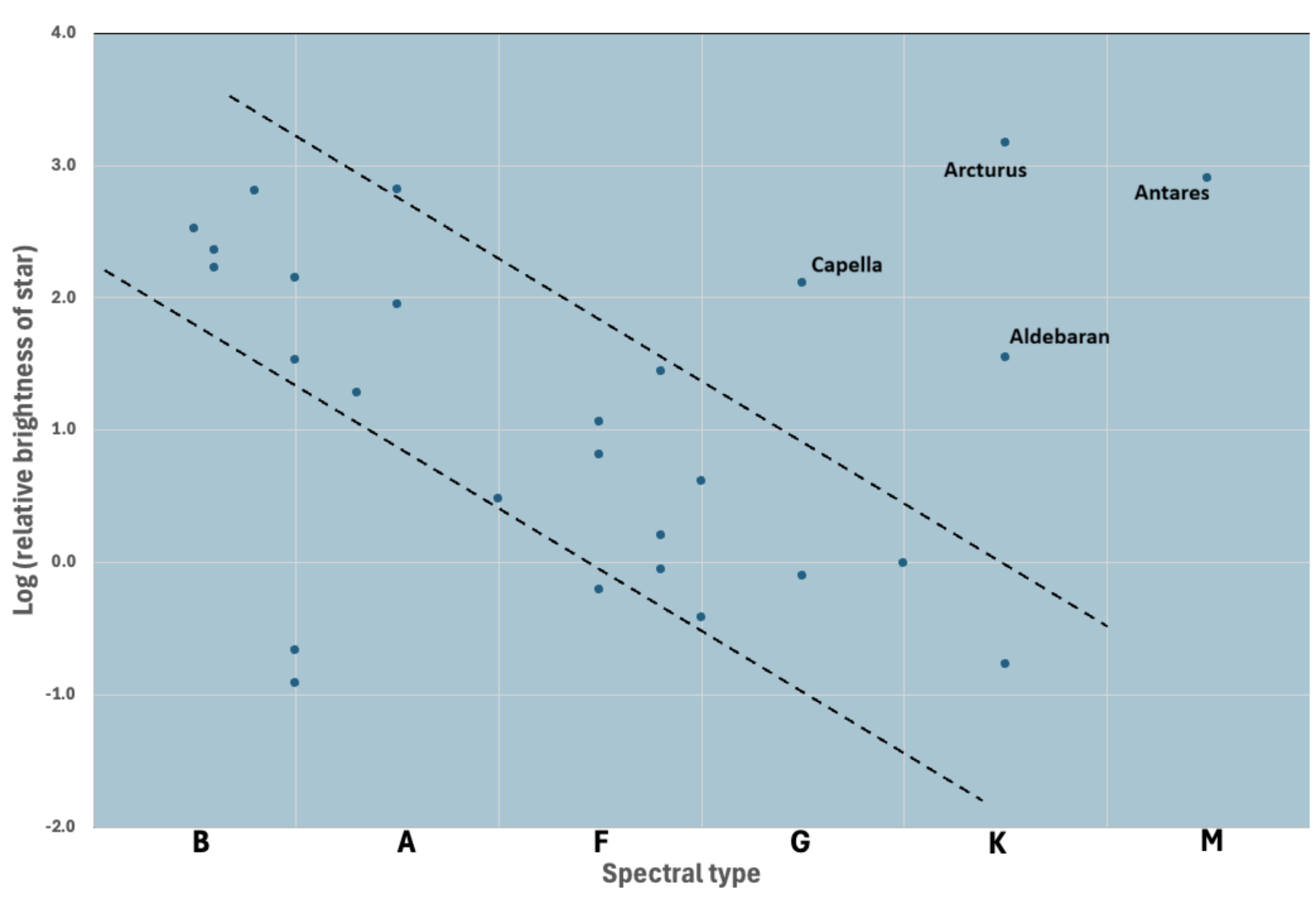

**Figure 4: Data from Gore's 1905 *Monthly Notices* paper 'On the Relative Brightness of Stars'**